\documentclass[traditabstract]{aa}
\usepackage{graphicx}
\usepackage{natbib}
\usepackage{lscape}
\usepackage{txfonts}

\begin{document}

\titlerunning{RR~Lyr variable stars in M~32} 
\authorrunning{G.~Fiorentino et al.}
\title{The ancient stellar population of M~32:\\
 RR~Lyr Variable stars confirmed} 

\author{G. Fiorentino\inst{1}, R. Contreras Ramos\inst{2}, E. Tolstoy\inst{3}, G. Clementini\inst{1}, A. Saha\inst{4}}
\authorrunning{G.~Fiorentino et al.~}
\institute{INAF-Osservatorio Astronomico di Bologna, via Ranzani 1, 40127, Bologna
\email{giuliana.fiorentino@oabo.inaf.it}
\and
Dipartimento di Astronomia, Universit\'{a} degli studi di Bologna,  via Ranzani 1, 40127, Bologna
\and
Kapteyn Astronomical Institute, University of Groningen, Groningen, The Netherlands
\and
NOAO, P.O. Box 26732, Tucson, AZ 85726, US}

   \date{Received June 9, 2011; accepted Dec 15, 2011}

\abstract{
Using archival multi--epoch ACS/WFC images in the F606W and
F814W filters of a resolved stellar field in Local Group dwarf
elliptical galaxy M~32 we have made an accurate Colour$-$Magnitude
Diagram and a careful search for RR~Lyr variable stars.  We identified
416 bona fide RR~Lyr stars over our field of view, and their
spatial distribution shows a rising number density towards the centre
of M~32. These new observations clearly confirm the tentative result
of Fiorentino et al. (2010), on a much smaller field of view,
associating an ancient population of RR~Lyr variables to M~32. We
associate at least 83 RR~Lyr stars in our field to M~32.

In addition the detection of 4 Anomalous Cepheids with masses in
the range 1.2$-$1.9~M$_{\odot}$ indicates the presence of relatively
young, 1$-$4~Gyr old, stars in this field. They are most likely
associated to the presence of the blue plume in the
Colour$-$Magnitude Diagram. However these young stars are unlikely to be
associated with M~32 because the radial distribution of the blue plume
does not follow
the M~32 density profile, and thus they are more likely to belong to
the underlying M~31 stellar population. Finally the detection of 3
Population II Cepheids in this field gives an independent measurement
of the distance modulus in good agreement with that obtained from
the RR~Lyr, $\mu_0$=24.33 $\pm$ 0.21 mag.

%

}


\maketitle
\section{Introduction}

The compact dwarf elliptical galaxy M~32 is one of the few elliptical
galaxies close enough that, with the help of HST, direct observations
can be made of its ancient resolved stellar population
\citep[e.g.,][]{grillmair96, alonsogarcia04, fiorentino10b,
monachesi11}.  Another example is the peculiar giant elliptical galaxy
Cen~A, at a distance of 3.8~Mpc, where the red giant branch (RGB) and
red clump (RC) have been resolved in the outer halo
\citep[e.g.,][]{rejkuba11}. More distant elliptical galaxies, and also
the extremely high surface brightness inner regions of M~32, can only
be studied in integrated light \citep[e.g.,][]{renzini06, rose94,
trager00, coelho09}.  Integrated light provides a picture of the mean
properties of a galaxy, but these analyses are rarely unique and
cannot avoid a strong bias from the dominant stellar population.

The search for traces of an old ($>$10~Gyr) stellar population 
is the main motivation for this study of the resolved stellar
population of M~32. RRLyr variable stars are
unequivocal indicators of the presence of a stellar population $>$
10~Gyr old.  There have been a number of detailed studies of the RRLyr
population of M~31, both with ground$-$based wide field surveys of the halo \citep{pritchet87,dolphin04}, and HST monitoring of M~31
globular clusters \citep{clementini09} and small halo and disk fields
\citep{brown04,jeffery11}. Thare have also been studies in the region around M~32
\citep{alonsogarcia04,sarajedini09,fiorentino10b}.

A recent analysis of the resolved stellar population of M~32 using the
ACS/HRC camera, which has a tiny field of view $\sim$ 30x30 arsec square
(see small blue square in Fig.~\ref{fig:fc}) but very high spatial resolution
(a factor two better than WFC/ACS), concluded that the bulk of the
stellar population has an age in the range $8 - 10$~Gyrs old, with a
mean metallicity, [Fe/H]$\sim$ $-$0.2 dex \citep{monachesi11}.
However their Colour-Magnitude-Diagram (CMD) did not reach the Main Sequence Turn Off stars of the
oldest stellar population. They also associated a faint blue plume (BP) of
stars in the CMD to $\geq$1~Gyr old stars in M~32.  Using the same
ACS/HRS data 17 RRLyr were found, and it was tentatively suggested
that $7_{-3}^{+4}$ could be associated to M~32 \citep{fiorentino10b}.
The small field of view made this result highly uncertain simply due
to the poor spatial sampling of a sparse population and the
significant contamination from M~31.  Here we use archival ACS/WFC
data, with significantly better spatial coverage (see large red
square in Fig.~\ref{fig:fc}), to revisit the
detection of RRLyr associated with M~32.

\section{Data reduction and analysis}\label{data}

\begin{table*}
\caption{Log of the M~32 ACS/WFC archival data. }. \label{table:data} 
\centering
\scriptsize
\begin{tabular}{cccccc}
Field&$\alpha_{J2000.0}$&$\delta_{J2000.0}$&Filter&Exposure time&Date\\
\hline
\hline
F1&00$^h$42$^m$56$^s$&+40$^{\circ}$50$^{\prime}$50$^{\prime\prime}$&$F606W$&8$\times$1000&24 Nov 2004\\
F1&00$^h$42$^m$56$^s$&+40$^{\circ}$50$^{\prime}$50$^{\prime\prime}$&$F606W$&5$\times$1000&25 Nov 2004\\
F1&00$^h$42$^m$56$^s$&+40$^{\circ}$50$^{\prime}$50$^{\prime\prime}$&$F606W$&5$\times$1000&10 Dec 2004\\
F1&00$^h$42$^m$56$^s$&+40$^{\circ}$50$^{\prime}$27$^{\prime\prime}$&$F814W$&8$\times$1580&24 Nov 2004\\
F1&00$^h$42$^m$56$^s$&+40$^{\circ}$50$^{\prime}$27$^{\prime\prime}$&$F814W$&5$\times$1580&25 Nov 2004\\
F1&00$^h$42$^m$56$^s$&+40$^{\circ}$50$^{\prime}$27$^{\prime\prime}$&$F814W$&5$\times$1580&10 Dec 2004\\
\hline 
\end{tabular}
\end{table*}
\normalsize

\begin{figure}
\includegraphics[width=8.5cm]{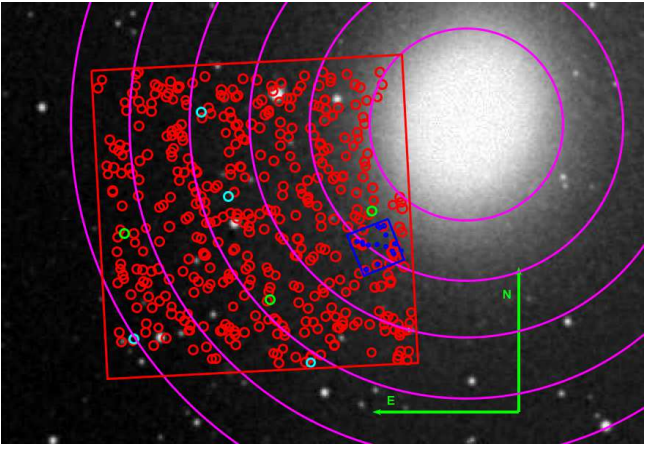}
\caption{A 6~x~5~arcmin DSS2 where our ACS/WFC field
is shown as a large (red) square. The HRC field with its RRLyr
\citep[from][]{fiorentino10b} is shown as a small (blue) square. The
RRLyr, Anomalous Cepheid and Population II Cepheid
variables detected in the ACS/WFC field are shown as red, cyan and
green points, respectively.  Circles centred on M~32 used in the density analysis are plotted in magenta. }\label{fig:fc}
\end{figure}

We used 36 archival ACS/WFC images\footnote{Proposal id: 9392; PI:
M. Mateo}, covering a 3.3 x 3.3 arcmin square region, centred at 00 42
56 $+$40 50 50 (J2000), and extending from 1 to 4.5~arcmin from the
centre of M~32, see Fig.~\ref{fig:fc} and Table~\ref{table:data}. 
The cadence of the images was chosen with the aim of picking up short
period variable stars (periods $<$1~day), which includes RRLyr. The
sensitivity of the temporal sampling to these short periods is close to $\sim$100\%
complete, for variable stars with periods $<$1~day.

\vspace{0.1cm}

\subsection{Photometric analysis}
We used the same data reduction and analysis techniques described in
\citet{fiorentino10a,fiorentino10b}.  The PSF$-$fitting photometry was
carried out using DOLPHOT, a version of HSTphot modified for ACS
images \citep{dolphin00a}.  DOLPHOT returns a time series of
measurements for each image as well as a final mean magnitude for all
stars found on all the individual frames.  DOLPHOT also makes
automatic aperture corrections, following the prescription of
\citet{sirianni05}, and corrects for charge transport efficiency (CTE)
effects \citep[as described in][]{dolphin00b}. Thus we end up with
a final catalogue consisting of a time series of 18 individual
photometric measurements in two filters, plus mean magnitudes, for
$\sim$400~000 stars over the entire ACS/WFC field of view.

There is clearly a very strong gradient in stellar density across our
field of view, which increases towards the centre of M~32. This
gradient leads to
large variations in the crowding of the stellar images. We thus also
performed artificial star tests to quantify the completeness of the
photometric catalogue over the field of view. A mean
 completeness above 90\% for the whole field is only found for stars
with m$_{F606W}$ $\leq$ $24.5$ mag. For stars fainter than this, the completeness drops rapidly towards the centre of M~32, as
the stellar density increases. This has to be carefully corrected
  before conclusions can be drawn about grandients across this
  field. We come back to this point in Section~\ref{sec:5}.

We also tested our photometric catalogue for evidence of differential
reddening across the field using the well established method of
\citet{piotto99}. We divided our field into 64 26x26 arcsec$^2$
regions and selected one of these as reference. Then for each sub-region we compared
the CMD properties with the reference. No evidence for differential reddening was found,
down to a limit of m$_{F606W}=26$ mag, and thus we adopted a single
reddening value of E(B$-$V) = 0.08 mag \citep[A$_{F606W}$=0.21 mag,
A$_{F814W}$=0.13 mag, see][]{fiorentino10b,bedin05}, in our analysis.

\subsection{Search for variability}
\begin{figure}
\centering
\includegraphics[width=8.5cm]{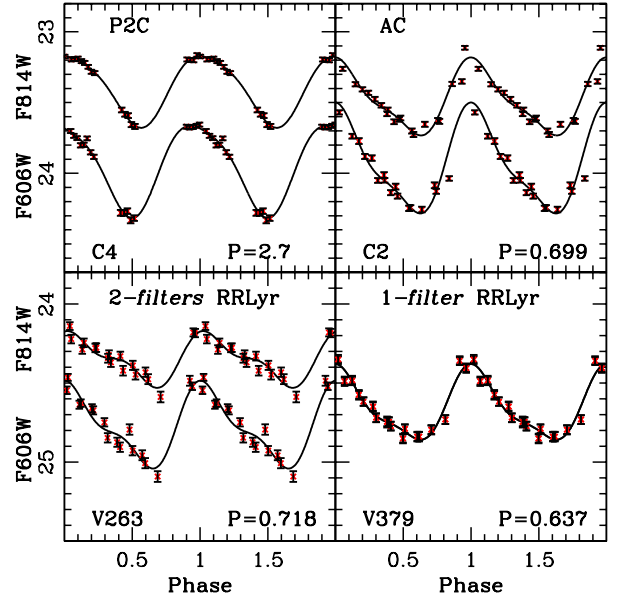}
\caption{An example of the quality of the light
  curves in F606W and F814W filters for four detected variable
  stars, a Population II Cepheid, an Anomalous Cepheid, a {\it
    one-filter} and a {\it two-filters} RRLyr. The ID, type and period have been labeled. The fitting model
  used to determine the mean magnitude and amplitude has been
  overplotted. The errorbars represent the photometric error returned
  by DOLPHOT.
}
\label{fig:lc}
\end{figure}

Details of the procedures to detect and classify variable stars are
described in \citet{fiorentino10b}.
 The average errors on the periods are $<$ 0.01 day, on the
  amplitudes of the variation are $<$ 0.1 mag and on the mean
  magnitudes are $<$ 0.1 mag. We identified 416 bona--fide
RRLyr.  We divided the RRLyr into two groups, {\it two-filters} (351
RRLyr) and {\it one-filter} (65 RRLyr).  For {\it two-filters}
candidates we were able to properly phase the F606W and F814W light
curves.  For the {\it one-filter} candidates we could estimate the
period using only one filter (usually F606W). We identified also 4 Anomalous Cepheids (AC) and 3 Population II Cepheids (P2C), see
Section~\ref{sec:4}.
In Fig.~\ref{fig:lc} we show the light curves we have obtained for 2
RRLyr, 1 AC and 1 P2C. We have also detected
61 variable candidates which cannot be accurately classified, they
include 27 possible RRLyr and 34 longer period variables.

We have recovered 9 of the 17 RRLyr ($\sim$ 53 \%) found in our
previous study with ACS/HRC \citep{fiorentino10b}, confirming the low
mean completeness ($\sim$ 57 \%) of the HB luminosity on the present
ACS/WFC images so close to M~32.

\section{The RRLyr variable stars}\label{sec:3}

\begin{figure}
\centering
\includegraphics[width=8.5cm]{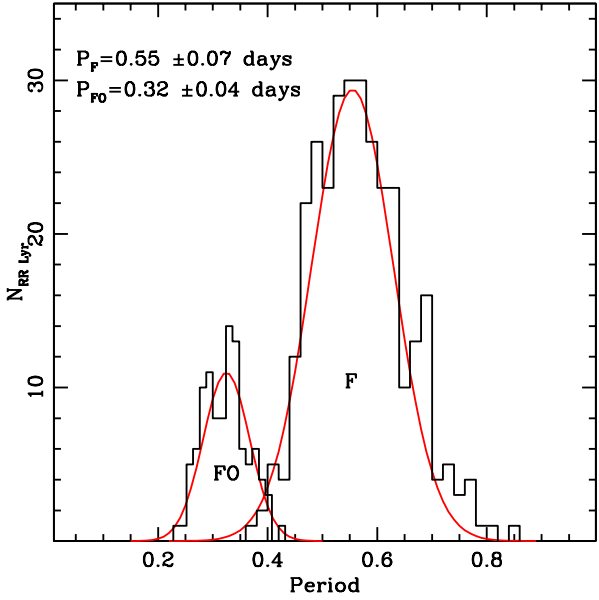}
\caption{
 The period distributions of the fundamental (right) and
  first-overtone (left) RRLyr. The periods corresponding to the
  gaussian (red) peaks are given as P$_F$ and P$_{FO}$.
}
\label{fig:per}
\end{figure}

From the 416 bona$-$ fide RRLyr we have detected, we classify 314
as fundamental pulsators (F or RR$_{ab}$) and 102 first$-$overtone
pulsators (FO or RR$_c$). Table~\ref{table:rr} lists the RRLyr along with their
properties such as position, period, pulsation type
classification, mean intensity$-$weighted magnitudes, amplitudes and de$-$reddened mean magnitudes transformed in the
Johnson$-$Cousin photometric system. The RR Lyrae have been named
using their increasing distance from the centre of M~32.

The mean period distribution of the F pulsators matches
a gaussian peaked at P$_F =$0.55 days with $\sigma = 0.07$ days, see
Fig.~\ref{fig:per}, which is consistent with Oosterhoff type I.
Similarly, a mean period of P$_{FO} =$ 0.32 days with $\sigma =$ 0.04
days is found for FO pulsators, which is also consistent with an
Oosterhoff type I. The ratio between the number of FO and the total sample, N$_{FO}$/N$_{TOT}$ $\sim$
0.25. Our results are thus in agreement with the M~31 study performed by
\citet{sarajedini09} of two ACS fields close to
M~32. Other deep HST/ACS observations of RRLyr M~31
include six fields of the halo, disk, and giant stellar stream have been
studied by \citet{jeffery11}. The RRLyr
of these fields appear to mostly be of Oosterhoff I type. Comparing our results with these six M~31 fields which sample different components of M~31, the
properties of our sample resembles the giant stream of M~31, where they found $<P_F>$=0.56 and
N$_{FO}$/N$_{TOT}$=0.33.

\begin{table*}
\caption{The RRLyr variable stars we have identified. The mean F606W
  and F814W magnitudes are intensity$-$weighted over the light
  curve. V$_0$ and I$_0$ are the de$-$reddened mean magnitudes in the
Johnson$-$Cousin photometric system obtained following the prescriptions by \citet{sirianni05}. Type classification, RR$_{ab}$ and RR$_c$, indicate fundamental and first$-$overtone
pulsation modes, respectively.} \label{table:rr} 
\centering
\scriptsize
\begin{tabular}{ccccccccccc}
ID&$\alpha_{J2000.0}$&$\delta_{J2000.0}$&Period&Type&$<$m$_{F606W}>$&A$_{F606W}$&$<$m$_{F814W}>$&A$_{F814W}$&$<$m$_{V_0}
>$&$<$m$_{I_0} >$\\
\hline
\hline
V1 & 0:43:01.7183 & 40:52:33.583 & 0.306 & RRc  & 25.080 & 0.379 & 24.799 & 0.209 & 24.910 & 24.661\\
V2 & 0:43:01.8445 & 40:52:32.564 & 0.458 & RRab & 25.459 & 1.095 & 25.118 & 0.667 & 25.304 & 24.980\\
V3 & 0:43:03.9260 & 40:52:29.779 & 0.406 & RRab & 24.983 & 0.522 & 24.627$^a$    & - & - & - \\			     
V4 & 0:42:50.5077 & 40:52:29.038 & 0.584 & RRab & 24.940 & 0.682 & 24.530 & 0.475 & 24.794 & 24.392\\
V5 & 0:42:57.7187 & 40:52:28.670 & 0.310 & RRc  & 25.422 & 0.460 & 25.082 & - & - & - \\
V6 & 0:42:51.6067 & 40:52:27.570 & 0.399 & RRc  & 24.846 & 0.451 & 24.465 & 0.287 & 24.703 & 24.326\\
V7 & 0:42:47.0945 & 40:52:27.376 & 0.690 & RRab & 25.182 & 1.073 & 24.151 & - & - & - \\
\hline 
\end{tabular}
\\
$^a$ for one$-$filter RRLyr we have only one
mean magnitude, thus we can not estimate the corresponding amplitude
or Johnson-Cousin\\
 magnitudes, we can only indicate the Dolphot mean magnitude.
\end{table*}
\normalsize

\subsection{Bailey diagram and metallicity}

\begin{figure}
\centering
\includegraphics[width=8.5cm]{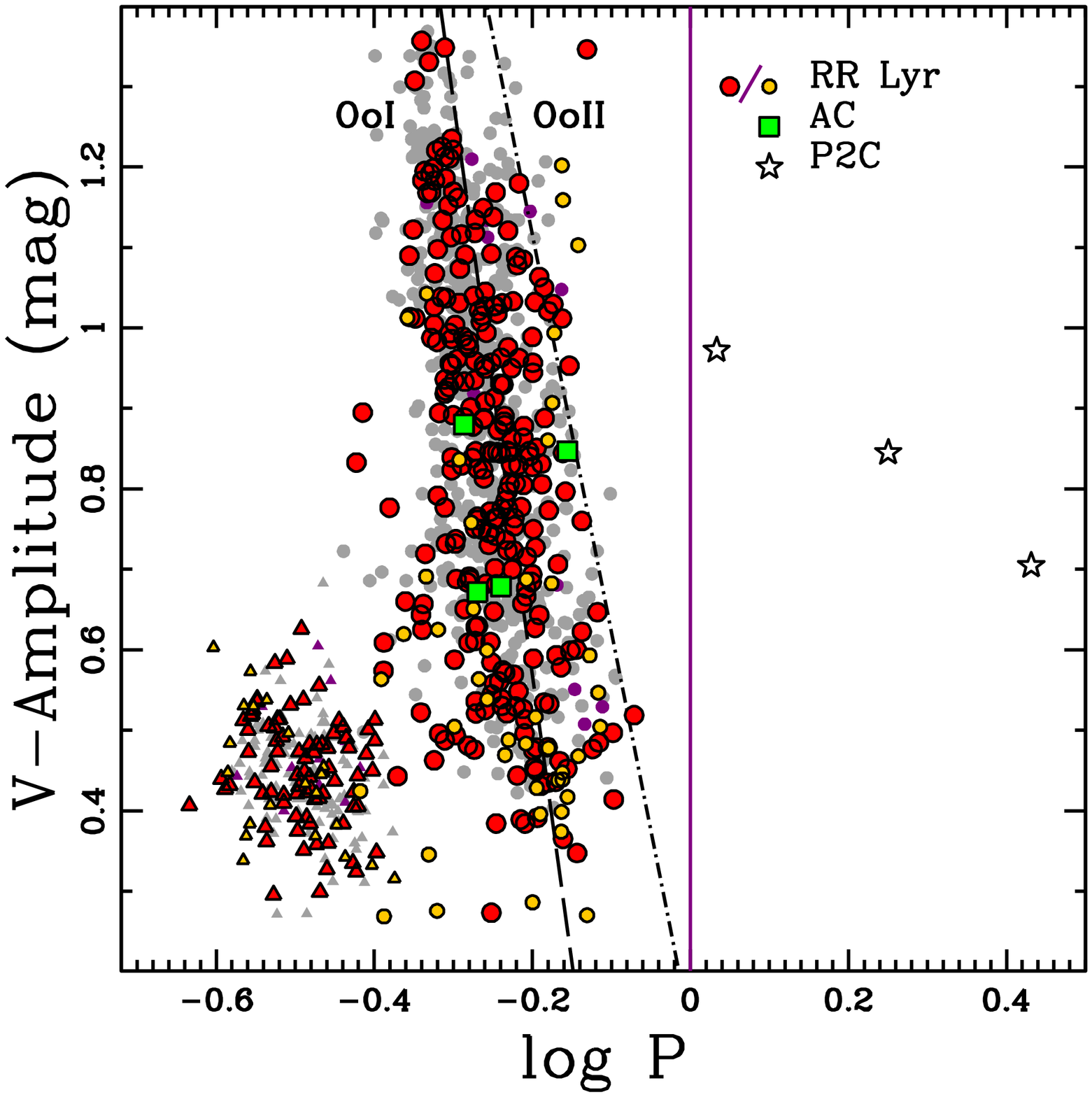}
\caption{
The V filter amplitude -- period (P) Bailey
diagram for our observations (coloured symbols) and the observations
of M~31 RRLyr  from
\citet{sarajedini09} and \citet{brown04} as grey and magenta symbols. The
Oosterhoff type loci, OoI (dashed) and OoII (dot$-$dashed), are shown
\citep[from][]{clement00}. For the RRLyr the triangles are first-overtone pulsators and the circles are fundamental
pulsators. RRLyr {\it two-filters} are plotted in red and only in {\it
  one-filter} in orange. Anomalous Cepheids and Population II Cepheids are indicated with squares (green) and stars,
respectively.
}
\label{fig:bailey}
\end{figure}

Fig.~\ref{fig:bailey} shows the Bailey diagram of the
variable stars detected in the present study overlaid on the M~31
studies of \citet{sarajedini09,brown04}. This diagram divides into two regions,
at log P=0, to distinguish the variable star types. At log
P$< 0$ are the RRLyr and AC and at log P$> 0$ the
P2C. The
${F606W}$ amplitudes have been transformed to V$-$filter amplitudes
\citep{brown04} using A$_{F606W}$=92~\% A$_V$ to directly compare the Oosterhoff (Oo) type loci \citep[see][and references
therein]{clement00}. In Fig.~\ref{fig:bailey} we see that F RRLyr are
predominantly consistent with an Oo type I, confirming the Oo classification
based on the periods alone.


\begin{figure}
\centering
\includegraphics[width=8.5cm]{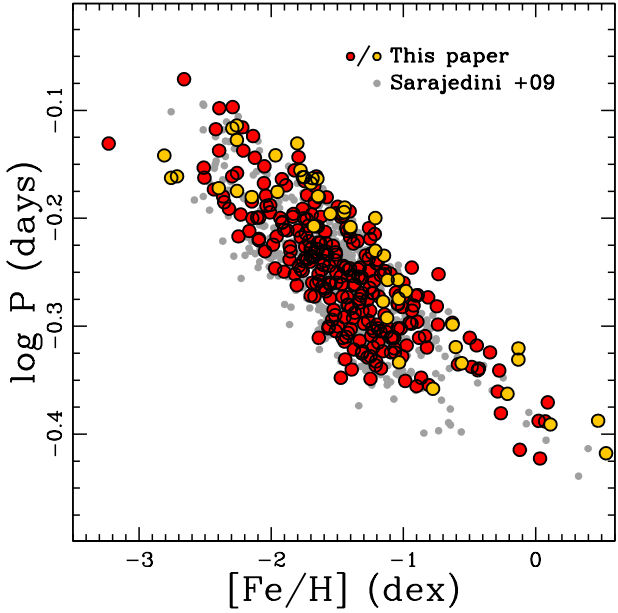}
\caption{
The metallicity ([Fe/H]) $-$period (P) distribution for our sampled fundamental RRLyr. The colour
code is the same used in Fig.~\ref{fig:bailey}. Undelying our
observation those os \citet{sarajedini09} are shown as grey symbols.
}
\label{fig:met}
\end{figure}


Next, we use this V amplitude$-$period relation 
\citep[given by][]{alcock00} to determine the individual metallicities
  of our entire sample of RRLyr. This is a method commonly used in the literature for
RRLyr studies without colour information. We found a mean
metallicity [Fe/H]=$-1.44\pm$0.55 dex for our sample, comparable with
what was found in M~31. 
 \citet{sarajedini09} found
[Fe/H] $= -1.45\pm$0.45 for their F1 field (the closest to M~32) and
[Fe/H] $= -1.54\pm$0.43 for their F2.  In
Fig.~\ref{fig:met} we show a comparison between individual
metallicities for our sample and Sarajedini's sample.  We can
see that our results are in very good agreement with their
studies. We also notice that in our field, and also in the Sarajedini
field closest to M~32, there are a few RRLyr of solar
metallicity. This means that we observe more
metal$-$rich RRLyr than have been found in the halo, disk and giant stream of M~31
by \citet{brown04} and \citet{jeffery11}, where using the same method,
no RRLyr with metallicity larger than [Fe/H] $>$$-1$~dex have been found.

\subsection{The Horizontal Branch}

\begin{figure*}
\centering
\includegraphics[width=8.5cm]{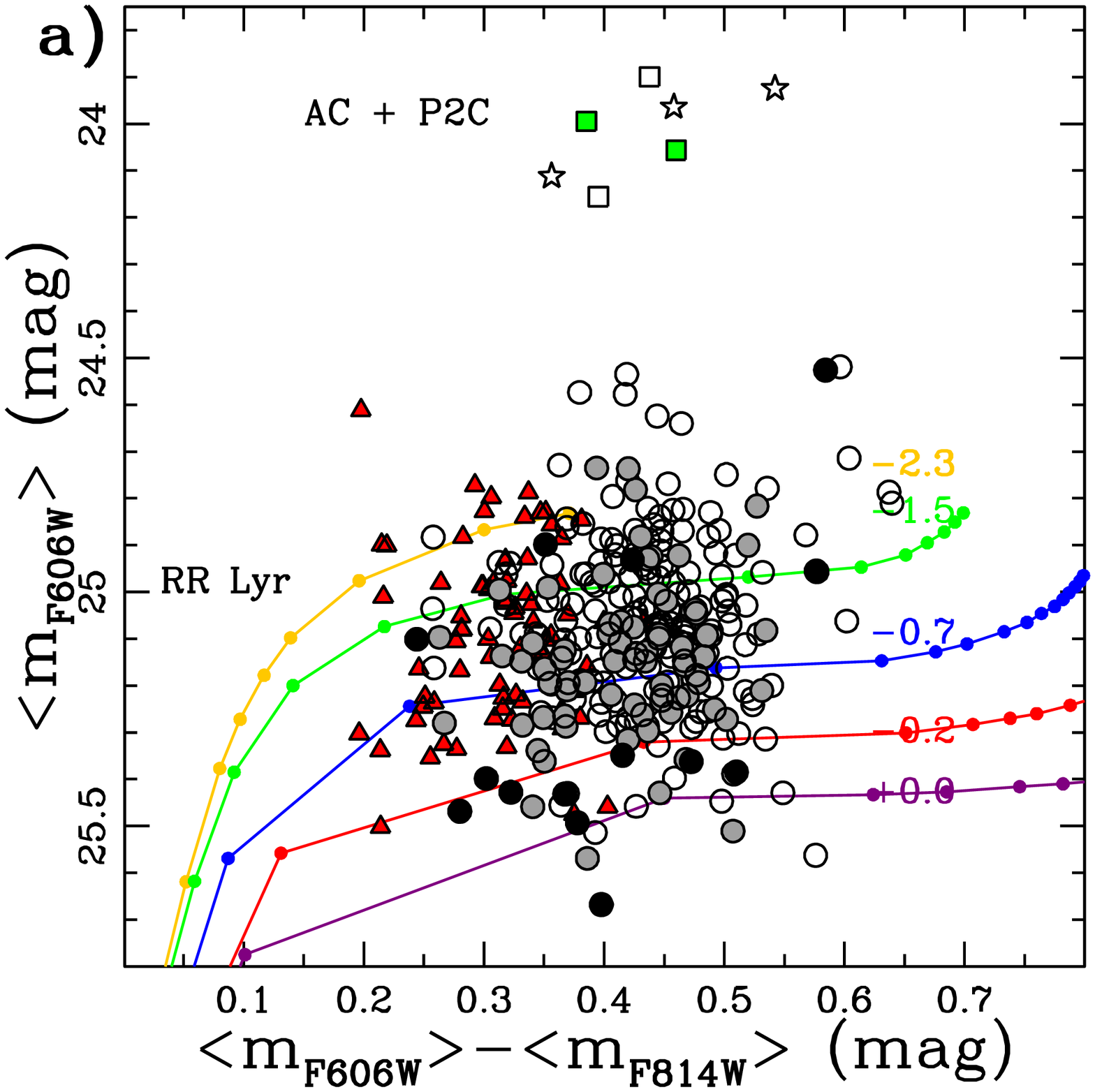}
\includegraphics[width=8.5cm]{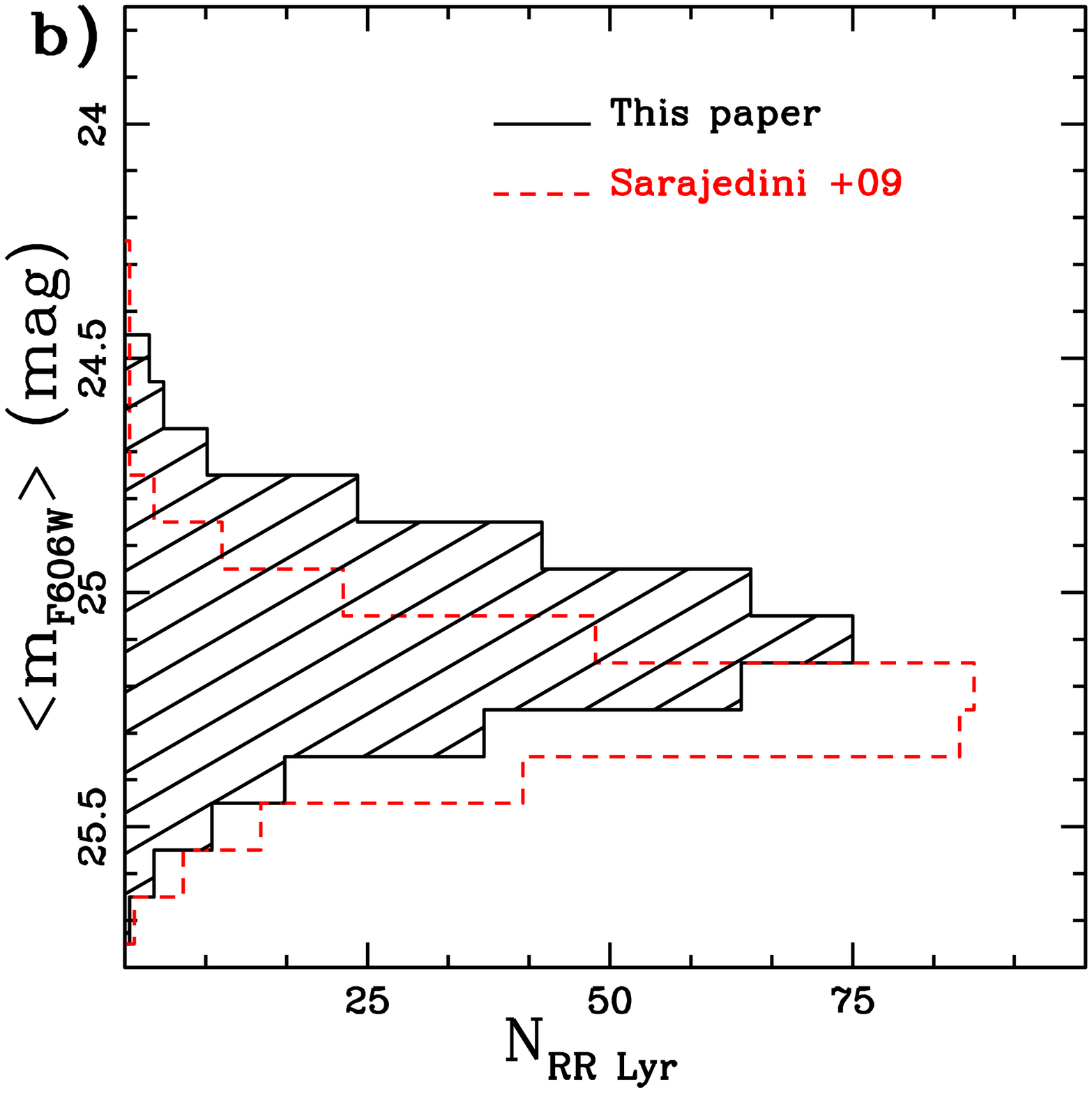}
\caption{
{\bf a):} the RRLyr region of the CMD, where only 
variable stars detected in two filters, with well defined colours and magnitudes, are shown. Red triangles are the same as in Fig. 2a, but the 
colouring of the F pulsators is determined by their metallicities, where
empty circles are metal$-$poor ([Fe/H]$\le-1.5$ dex), grey circles are metal$-$intermediate ($-0.5\geq$[Fe/H]$\geq-1.5$ dex) and filled circles 
are metal$-$rich ([Fe/H]$\geq-1$ dex). AC and P2C are indicated with squares and stars,
respectively. AC pulsating in first$-$overtone mode are highlighted
with filled colour (green). The coloured lines are HB tracks from \citep{pietrinferni04} for different
metallicities, as labeled.
{\bf b):} the luminosity function for our 2-{\it filter} RRLyr sample (black solid line) compared with that found by
\citet{sarajedini09} (red dashed line).
}\label{fig:hb_f606w}
\end{figure*}

The Horizontal Branch (HB) is almost invisible in our CMD,
because it is sparsely populated and dominated by contamination from other
populations, such as an extensive and broad RGB and a young BP
of stars in the main sequence region. The only way to reliably
identify the HB is
using RRLyr. In Fig.~\ref{fig:hb_f606w}a
we show the CMD of the Horizontal Branch region for our {\it
two-filters} RRLyr, for which we have light$-$curve weighted mean
magnitudes and colours. The RRLyr define the HB and thus give us the opportunity to
determine the mean distance modulus of the sample. To derive
the distance modulus, we transformed mean m$_{F606W}$ and m$_{F814W}$
(columns 8 and 10 in Table~\ref{table:rr}) into Johnson$-$Cousins m$_V$
and m$_I$ (columns 12 and 13 in Table~\ref{table:rr}). We thus obtained a de$-$reddened V magnitude of the HB, $<V_0>$ = 24.95  $\pm$ 0.18 mag. Then, assuming the average
metallicity found above and using the M$_V -$[Fe/H] relationship
\citep{carretta00}, we obtain a distance modulus of $\mu_0 =$24.33
$\pm$ 0.21 mag. The large uncertainty is due to both the large
metallicity spread (see Fig.~\ref{fig:met}) and the large scatter on
the HB ($\pm 0.5$~mag), as seen in Fig.~\ref{fig:hb_f606w}a.

In Fig.~\ref{fig:hb_f606w}b, we show the luminosity histogram of
our {\it two-filters} RRLyr and we compare it with that for field F1
from \citet{sarajedini09}. The peak of our luminosity distribution is
brighter than Sarajedini's by $\sim$0.16 mag. This could be due to
the significantly different surface brightness of the two fields,
which is $\sim $22 mag/arcsec$^2$ in our field and $\sim $25
mag/arcsec$^2$ in Sarajedini's. A brighter surface brightness implies
more severe blending effects that can lead to brighter magnitudes for
individual stars \citep[e.g., ][]{fiorentino10b}. However, the broad
spread we find in m$_{F606W}$ does not change using only the least
blended\footnote{as defined by the DOLPHOT crowding parameter.} 
RRLyr in our sample. This suggests that
this spread is not likely to be caused by photometric errors. The spread could
be due to either the RRLyr being at a range of different distances
and/or having a large range in metallicity. Note that in both
cases differences in the evolutionary states may affect the RRLyr
luminosity, which would further increase the final HB
luminosity range. In
Fig.~\ref{fig:hb_f606w}a we show the effect of a metallicity range
using the theoretical HB tracks for a range of
metallicity at the mean distance modulus derived above. The
metallicity spread inferred for the RRLyr is consistent
with the large range derived from the periods and amplitudes in
Fig.~\ref{fig:bailey}. However this does not account for the
whole HB spread. In addition, a distance spread is also very
likely because the M~32 field is mixed with M~31 population. Assuming the mean metallicity found above and moving the
relative zero age HB locus to fit the brightest and the faintest RRLyr in
Fig.~\ref{fig:hb_f606w}a, we find a distance modulus of
24.33$^{+0.47}_{-0.43}$ mag (or d = 734$^{+178}_{-131}$
Kpc).

To determine how many, if any, of the RRLyr in our sample are likely to be associated
with M~32, we need to look at their radial distribution from the centre of M~32.

\section{The Cepheid variable stars}\label{sec:4}

\begin{table*}
\caption{The confirmed Cepheid
  variable stars. Same as in Table~\ref{table:rr}}. \label{table:cep} 
\centering
\scriptsize
\begin{tabular}{ccccccccccc}
ID&$\alpha_{J2000.0}$&$\delta_{J2000.0}$&Period&Type& $<$m$_{F606W}>$&A$_{F606W}$&$<$m$_{F814W}>$&A$_{F814W}$&$<$m$_{V_0}
>$&$<$m$_{I_0}$$>$\\
\hline
\hline
C1 & 0:42:47.8102 & 40:50:53.238 & 1.78  & P2C     & 23.964 & 0.782 & 23.506 & 0.579 & 23.839 & 23.368 \\
C2 & 0:42:56.4848 & 40:51:06.308 & 0.70  & AC$-$F  & 23.899 & 0.784 & 23.462 & 0.551 & 23.765 & 23.323  \\
C3 & 0:42:57.9929 & 40:52:04.630 & 0.53  & AC$-$FO & 23.995 & 0.622 & 23.610 & 0.421 & 23.853 & 23.471  \\
C4 & 0:42:54.1987 & 40:49:53.248 & 2.70  & P2C     & 23.925 & 0.652 & 23.382 & 0.506 & 23.835 & 23.244  \\
C5 & 0:43:02.8871 & 40:50:43.142 & 1.08  & P2C     & 24.112 & 0.900 & 23.756 & 0.627 & 23.962 & 23.618   \\
C6 & 0:42:51.8780 & 40:49:07.727 & 0.57  & AC$-$FO & 24.057 & 0.629 & 23.597 & 0.412 & 23.909 & 23.448  \\
C7 & 0:43:02.5661 & 40:49:28.127 & 0.51  & AC$-$F  & 24.156 & 0.815 & 23.761 & 0.547 & 24.017 & 23.622 \\
\hline 
\end{tabular}
\end{table*}

\normalsize

In our sample we have identified 7 variable stars (described in Table~\ref{table:cep}) with luminosities
significantly higher than the HB, as shown in
Fig.~\ref{fig:hb_f606w}. These
stars are at luminosities and colours consistent with both AC and P2C
\citep{fiorentino06,dicriscienzo07}. The AC are core-He burning
stars of intermediate age ($\sim$1$-$6~Gyr old), with mass of 1$-$2M$_{\odot}$, and they cover a period range from a few hours to
$\sim$2~days. The P2C are lower mass stars ($\leq$0.8M$_{\odot}$) that
just finished their core$-$He burning phase and come from the blue
HB. They are very old ($>$10~Gyr) and typically have periods $>$1~day. Those 
with periods $<$3~days, are the so$-$called BL Hercules variable stars and they come from the {\it warm}
part of the HB \citep[e.g.][]{marconi11}.

\begin{figure}
\centering
\includegraphics[width=8.5cm]{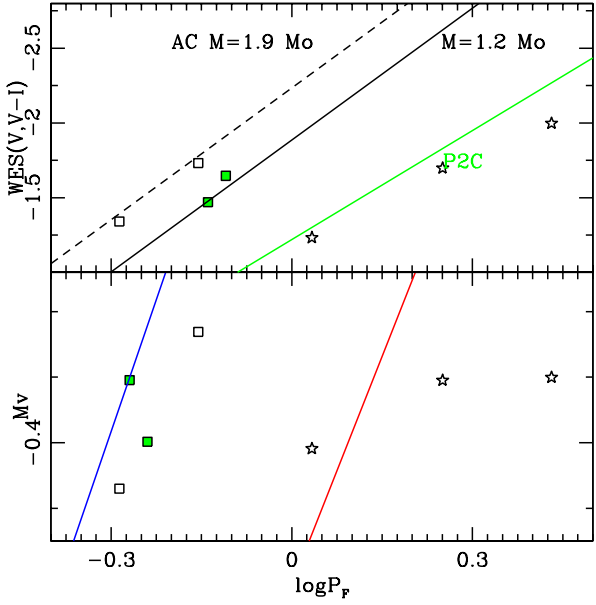}
\caption{
{\it Bottom:} 
Anomalous Cepheids (squares) and Population II Cepheids (stars) in the M$_V -$period (P) plane. 
First$-$overtone AC have been highlighted with filled colour (green). It is not
possible to make a clear classification between these two class of
variables. We have plotted the cold (red) and the warm (blue) edges of the pulsation instability strip as derived
by theoretical models \citep{fiorentino06}.
{\it Top:} Wesenheit plane, where the separation between the Anomalous
Cepheids and Population II Cepheids is clearer. The green solid
line represents the relation for Population II Cepheids, and the black solid and
dashed line represent the mass dependent Wesenheit relations for
Anomalous Cepheids, with
masses 1.2~M$_{\odot}$ and 1.9~M$_{\odot}$ respectively.
The distance modulus assumed is 24.33, as we found from RRLyr.
}
\label{fig:ac}
\end{figure}

The overlap in the period ranges of
AC and P2C makes a 
proper distinction between them complicated. 
We have plotted them all in the
Bailey diagram, see Fig.~\ref{fig:bailey}, where we see that
three of them
occupy the region with periods $>$1 day, and the remaining four lie
on the RRLyr Period$-$Amplitude relation. In Fig.~\ref{fig:ac}
(lower panel) we plot the pulsation Instability Strip derived from 
theoretical models \citep{marconi04} for AC, assuming the distance modulus derived in the previous section ($\mu_0$=24.33). 
There is only one variable not within the instability strip, and this corresponds to the 
longest period, namely C4. To check that the Cepheids within the
instability strip are 
genuine AC, we also use the colour information plotting them
in the reddening$-$free Wesenheit plane, i.e. WES(V,
V$-$I)=V$-$2.54E(V$-$I) see top panel of Fig.~\ref{fig:ac}. In this plane the
separation between these two classes of variables is very clear. 
The P2C follow a well
established Period$-$Luminosity relation, whereas the AC follow a more 
spread out Period$-$Luminosity relation that depends
on stellar mass. In Fig.~\ref{fig:ac} (top panel) we also show the theoretical Wesenheit
relations \citep{fiorentino07,dicriscienzo07} for
both AC and P2C. Thus we can classify three Cepheids (C4, C6, C7) as
P2C and the other four as AC (see Table~\ref{table:cep}). 

We can constrain the masses for the four AC, using our theoretical approach
\citep[extensively described in][]{marconi04,caputo05,fiorentino06} which
allow us to simultaneusly derive the mass and the pulsation mode of an
AC using the pulsation properties. We use the well defined mass
dependent Period$-$Magnitude$-$Amplitude relation, which is valid
only for F pulsator, and the Period$-$Magnitude$-$Colour
relations which are available for both
F and FO pulsators. Applying these relations to our four AC we
classify two of them F pulsators and the remaining FO pulsators (see Table~\ref{table:cep}) with masses in the range 1.2 $-$ 1.9
M$_{\odot}$. Their masses indicate these stars come from a stellar population
with ages between 1 and 4 Gyr old.

We notice that the distance modulus found from RRLyr agrees well with that from the P2C Wesenheit
relation. The only point of concern is the low luminosity of the
longest period P2C, C4. However mean luminosities and amplitudes could be
affected by the small temporal coverage used to sample the full pulsation cycle for these relatively long period variable stars.

\section{Radial distributions of Stellar Populations}\label{sec:5}

The stellar populations of M~31 and M~32 are so similar and mixed, in our field,
that the characteristics of M~32 can only be defined by looking for
features that follow the strong concentration of stars towards the
centre of M~32. This has to be done with extreme care, because
the crowding properties of the images are also changing closer to the
centre of M~32. We are mainly concerned to determine
if there is a RRLyr population that can be reliably
identified with M~32. To this end it is also useful to look at the
radial distribution of other stellar populations that can be
identified in our CMD.

\begin{figure}
\centering
\includegraphics[width=8.5cm]{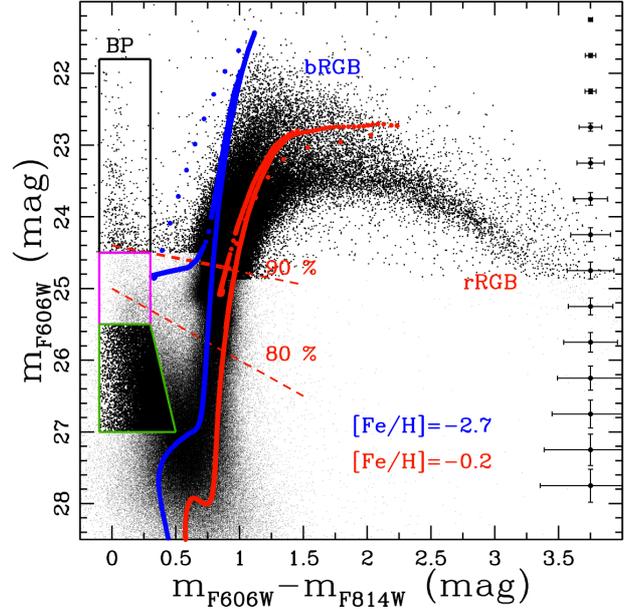}
\caption{
The ACS/WFC CMD, for all the detected stars. The mean
completeness levels (red dashed lines) and the photometric errors are shown. Over$-$plotted are 8 ~Gyr old reddened isochrones, from
\citet{pietrinferni04} with [Fe/H]$=-2.7$ dex (blue) and $-0.2$ dex
(red). Distance modulus and the reddening used are $\mu_0$=24.33 and E(B$-$V)=0.08 mag. Isochrones
and boxes have been used to select the populations. The blue (bRGB)
and red RGB (rRGB) on either side of [Fe/H]$=-0.2$ dex isochrone and the blue
plume (BP) which is separated into bright (black box), intemediate (magenta box) and faint (dark green
box) components.
}

\label{fig:cmd}
\end{figure}

In Fig.~\ref{fig:cmd} we show the CMD for all the detected stars in our ACS/WFC field of view. 
The two over$-$plotted isochrones are 8 Gyr old, which is the mean age for the bulk of the stellar population \citep{monachesi11}. 
To look for spatial gradients over our field towards the centre of
M~32, we select different stellar populations from the CMD, as
follows:\\ \\
{\it i)} the bright BP, which are young $\le$500~Myr
old main sequence stars;\\ 
{\it ii)} an intermediate BP, which includes a contribution from young main sequence
  stars ($\le$1~Gyr old) and also the blue HB;\\
{\it iii)} the faint BP, which is made up of main
  sequence stars $\le$4~Gyr old;\\ 
{\it iv)} RRLyr;\\
{\it v)} the blue RGB, which is those stars with m$_{F606W}\le$24.9 mag
that lie on the RGB between the blue and the red isochrones
and they should be mostly old metal$-$poor stars (-2.7$\le$[Fe/H]$\le$-0.2), but could also
include younger ($<8$~Gyr old) metal$-$rich stars ([Fe/H]$\le$-0.2);\\ 
{\it vi)} the red RGB, which is those stars with m$_{F606W}\le$24.9 mag
and on the red side of the red RGB and are all metal$-$rich RGB stars.\\

\begin{figure}
\centering
\includegraphics[width=8.5cm]{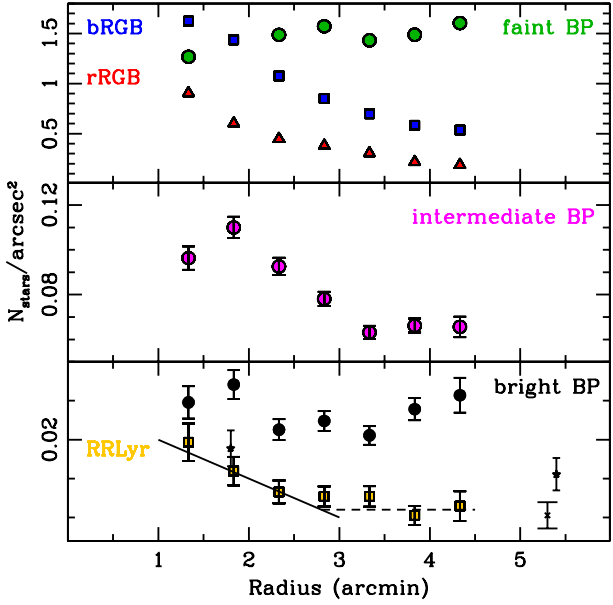}
\caption{
  The variation of the fraction of stars with the distance from the centre of M~32 for the populations
  selected from the CMD. The colour code is the same used in
  Fig.~\ref{fig:cmd}. The poissonian errorbars are
  plotted where they are larger than the symbols. Top panel, the
  bRGB, the rRGB and the faint BP have been plotted with circles, squares and stars, respectively. In the middle
  panel the intermediate BP distribution is plotted. The bottom
  panel shows the RRLyr and the bright BP radial distributions with
  squares and circles, respectively. The bright BP circles have been shifted
  of +0.01, to make the plot more clear. A comparison
  with previous RRLyr detections is also shown with stars \citep[F1
  and F2][]{fiorentino10b} and cross \citep[F1
  from][]{sarajedini09}. Solid and dashed lines represent the fit to
  M~32 and M~31 profiles, respectively.
}\label{fig:lf}
\end{figure}

These populations have been chosen to trace different ages and
  metallicities in the CMD attempting to separate different
  contributions coming from M~31 and/or M~32. In fact, we do not expect that young stars with ages less than 0.5~Gyr
belong to an elliptical galaxy such as M~32. On the other hand, we
expect that the contribution to the metal$-$rich component, as the red
RGB, is mostly coming from M~32, being almost invisible in other
accurate HST CMDs of the disk/halo components of M~31 \citep[e.g.][and reference therein]{brown06}.

We expect that the completeness will vary along with the distance from the centre of M~32. For each selected
population we have computed the completeness in annuli around M32, as
defined in Fig~\ref{fig:fc}. To account for the large colour baseline of our CMD,
we computed the completeness for all the blue stars (V$-$I less than
$\sim$ 0.7 mag) and for all the red stars (V$-$I larger than $\sim$ 0.7 mag) independently. As an
example, at the luminosity of the RRLyr on the HB
(m$_{F606W} \sim$ 25.07 mag) the mean completeness for RRLyr is
 57\% in the inner annulus (r$\sim$ 1.3 arcmin), whereas it
increases to 85\% in the outer annulus (r$\sim$ 4.3 arcmin).

 In Fig.~\ref{fig:lf}, we show the number of stars per arcsec$^2$ of
the selected stellar populations corrected for the completeness, as
determined by artificial star tests. Four out of the six selected
  populations, i.e. the blue and red RGB, the intermediate BP and the
  RRLyr, increases going to the centre of M~32 defining the radial profile of the elliptical galaxy. On the contrary, the faint and bright BPs
 (top panel) show, if anything, a shallow opposite trend, suggesting an association to M~31. 
Stars so young are a clear feature of the M~31
 field from other ACS/WFC studies, and this suggests that all the young
 stars observed in our field are likely associated to  
 M~31. However, the number statistics are low and it has
 been suggested by \citet{monachesi11} that there is a 
 BP component in M~32 (with ages $\geq$1~Gyr). 
We do not confirm this finding, however it could easily be hidden in 
 the huge contamination from M~31. We note that, among the BP
   components only the intermediate BP, which includes also stars from
   the HB, increases going towards the centre of M~32 (middle
 panel in Fig.~\ref{fig:lf}) then supporting the presence of the HB in
 M~32. If there were young stars in M~32 then also the other
BP components should show the same signature, but they do not.

Finally, the RRLyr distribution found in the
 present study shows an increasing trend towards the centre of M~32 and confirms previous results
 \citep{alonsogarcia04,fiorentino10b}. In Fig.~\ref{fig:lf} we also
 showed the RRLyr detections found from \citet{sarajedini09} and
 \citet{fiorentino10b}, which are in very good agreement with the
 RRLyr radial profile found in this paper. Our previous RRLyr detections, two points from \citet{fiorentino10b}, were uncertain because of 
their large poissonian errors due to the very small FoV
(30$x$30 arcsec) of ACS/HRC.
Our new detection of 416 RRLyr and their spatial variation over the
3.3$^{\prime} x$3.3$^{\prime}$ ACS/WFC field of view 
is the strongest evidence obtained so far for an ancient population in M~32. 

We also notice that the RRLyr distribution is quite flat going from
  outside to the centre of M~32 up to about 3 arcmin, suggesting that
  what we are observing for radii larger than  3 arcmin is a high
  background from M~31. Something similar can be seen also in the
  intermediate BP distribution supporting the idea of the strong M~31 background.

We can attempt an estimation of the fraction of the RRLyr that
belong to M~32 by assuming that the RRLyr outside $\sim$3 arcmin from the centre
of M~32 represent the M~31 background (N$_{stars}$/arcsec$^2$=0.011,
dashed line in Fig.~\ref{fig:lf}). Then, we can fit a slope to the
remaining observations ($r \leq$ 3 arcmin),
N$_{stars}$/arcsec$^2$=0.025$-$0.005*r (solid line in
Fig.~\ref{fig:lf}). The total number of
  stars whithin 3 arcmin from the centre to M~32 is 327 RRLyr, which
  includes the background from M~31. Then we can estimate the number
  of RRLyr inside the 3 arcmin, subtracting the contribution expected
  from the background of M~31. At least $\sim$~83 RRLyr could to be associated to M~32.

\section{Conclusions}

The radial density of RRLyr in our HST/WFC sample
clearly increases towards the centre of M~32. This convincingly shows 
that we have identified an ancient stellar population of at least 83 RRLyr
associated to M~32. These new RRLyr detections are in agreement with our
previous results \citep{fiorentino10b}, which is actually a bit lucky given the
statistics of the small field of view of HST/HRC.

The same trend towards the centre of M~32
is also followed by the other stellar components. Such as the blue and the red component of
the RGB. This suggests that M~32 contains both a moderately
metal$-$poor (-2.7$\le$[Fe/H]$\le$-0.2)
and a metal$-$rich ([Fe/H]$\ge$-0.2) stellar population. 

On the other hand, the bright and the faint blue plume of young main
sequence stars ($\le$ 4~Gyr old) has a more smooth distribution suggesting that it
is associated exclusively with the M~31 background. This result suggests that the 4 AC,
with masses from 1.2 to 1.9~M$_{\odot}$, found in our analysis are
likely to be associated to M~31.

We find compelling evidence for a large spread in the metallicities
of the RRLyr in our sample. They span -2.4$\le$[Fe/H]$\le$0 dex with a
mean metallicity of $<$[Fe/H]$> \sim$ -1.44$\pm$0.55
dex.

The RRLyr in this study are
classified as Oosterhoff type I, in agreement with recent results
\citep{brown04,jeffery11} for RRLyr observed in M~31. The properties
of our RRLyr sample resemble those of the stream of M~31. This could
suggest that M~32 could have contributed to the M~31 stream with its
old stellar population, or that the stream dominates M~31 at the
position of M~32.

\section*{Acknowledgments}
We thank A. Dolphin and D. Weisz  for their technical support running
DOLPHOT on many WFC/ACS images simultaneously. We are grateful to
A. Mucciarelli for his help in calculating the reddening map of our
WFC/ACS field. We thank S. Cassisi, M. Di Criscienzo, G. Greco,
M. Bellazzini and G. Bono for useful discussions. GF particularly
thanks R. Merighi his technical support. The authors acknowledge
  an anonymous referee for his/her suggestions that improved the
  content and the readability of the manuscript. GF and ET have been supported by a NWO$-$VICI grant and GF by the INAF fellowship 2009 grant. ET thanks the hospitality of Observatiore de la C\^{o}te d'Azur.
\bibliographystyle{aa}


\end{document}